\documentclass{article}

\usepackage{PRIMEarxiv}

\usepackage[utf8]{inputenc} 
\usepackage[T1]{fontenc}    
\usepackage{hyperref}       
\usepackage{url}            
\usepackage{booktabs}       
\usepackage{amsmath,amsfonts,amssymb}
\usepackage{nicefrac}
\usepackage{microtype}
\usepackage{fancyhdr}
\usepackage{graphicx}
\usepackage{array}
\usepackage{siunitx}
\usepackage{pifont}
\usepackage[table]{xcolor}
\usepackage{algorithm}
\usepackage{algorithmic}
\graphicspath{{media/}}

\usepackage{tikz}
\usetikzlibrary{arrows.meta, shapes.geometric, positioning}

\title{Discovering Explicit Magnetic Core Loss Equations via \\ Learnable Symbolic Sparse Identification
}

\author{
  Haoyu Wang$^{1}$, \quad
  Jialin Zheng$^{2}$\thanks{Corresponding author.}, \quad
  Yihao Wu$^{1}$, \quad
  Ziyang Xu$^{1}$, \quad
  Alex Hanson$^{1}$ \\[6pt]
  $^{1}$Department of Electrical and Computer Engineering,
  The University of Texas at Austin, Austin, TX 78712, USA \\[6pt]
  $^{2}$Department of Electrical and Computer Engineering,
  Princeton University, Princeton, NJ 08544, USA \\[6pt]
  \texttt{\{wanghaoyu, yw25243, ziyangxu, ajhanson\}@utexas.edu} \quad
  \texttt{jz8197@princeton.edu}
}

\begin{document}
\maketitle

\vspace{-0.5 cm}
\begin{abstract}
Explicit magnetic core loss equations with simple expressions and physical interpretability are significant tools in the design of high-frequency power magnetics. Traditional fits to empirical data like the Steinmetz Equation (SE) often struggle with accuracy, whereas modern machine learning approaches improve precision but deviate from physics. To fill this gap, this paper proposes a Learnable Symbolic Sparse Identification (LSSI) framework for data-driven equation discovery. Specifically, LSSI reformulates magnetic core loss equations for sinusoidal drives as a symbolic regression problem derived directly from experimental data. Building upon the SE, an expanded library of candidate functions are introduced and a sparse identification framework is implemented to select the dominant ones. More importantly, crucial parameters like exponents and coefficients of candidate functions are treated as learnable ones, simultaneously achieving equation simplicity and high expressiveness of the underlying fractional power laws.
Experimental results demonstrate that LSSI achieves superior accuracy with a state-of-the-art $\mathbf{R^2}$ of $\mathbf{0.9999}$ and a MAPE of $\mathbf{1.04\%}$ through a highly compact explicit equation containing only $\mathbf{4}$ active terms. Furthermore, it drastically reduces the parameter count from $\mathbf{4417}$ in neural network methods to $\mathbf{15}$, showcasing exceptional compactness and efficiency. The LSSI framework thus provides a physically transparent and highly accurate solution suitable for complex modern magnetic characterization and design.
\end{abstract}

\vspace{0.5 cm}


\section{Introduction}

The continuous push toward higher power densities in power electronics has significantly elevated the operating frequencies of magnetic components into megahertz domains \cite{7370781, 9893537, 11516849, 10132875, 10242261}. Consequently, magnetic core losses have become a dominant factor in the overall efficiency and thermal limits of modern power converters. However, high-fidelity modeling of volumetric core loss across a broad spectrum of frequencies and magnetic flux densities remains a fundamental challenge, as the underlying physical mechanism exhibit highly nonlinear dependencies under diverse operating conditions.

While establishing accurate closed-form analytical equations for core loss is highly attractive for magnetic design, it is challenging due to the highly nonlinear physics. Traditional empirical power laws, such as the Steinmetz Equation (SE) and its variants, offer simplicity but suffer from large prediction errors due to their fixed structures \cite{1457110, 5955126}. Recently, machine learning (ML) methods including traditional Random Forest (RF) \cite{10509368} and advanced Neural Networks (NNs) (e.g., Feedforward NN (FNN) \cite{10502151, 11589405} and Physics-Informed NN \cite{11358392, 11366028, 11126536}) have emerged to map these complex nonlinearities with great accuracy. However, these methods act as mathematical black boxes that are entirely ignorant of physical laws and yield no analytical insights, thus losing the speed and design intuition that explicit equations offer in practical magnetic design.

To restore mathematical transparency and design convenience, Symbolic Regression (SR) has been introduced to automatically discover explicit equations in nonlinear systems \cite{9180100, 11106929, 11592694}. Nevertheless, standard SR methods typically search over a large number of library candidate functions and struggle to achieve sparsity, whereas magnetic loss is generally dominated by only a few physical terms. Moreover, the candidate functions are usually built from predefined fixed parameters (e.g., discrete integer exponents), whereas magnetic loss inherently follows fractional power laws, as evidenced by the non-integer exponents in the classical SE. Forcing fixed integer bases to approximate fractional scaling inevitably leads to overly complex expressions that are both inaccurate and unfavorable.

To bridge the gap between ML methods and physical laws, this paper proposes a Learnable Symbolic Sparse Identification (LSSI) framework to discover the explicit magnetic core loss equations directly from experimental data. The framework establishes a candidate library with learnable parameters and proposes a joint optimization scheme to isolate the dominant terms and precise parameters, thereby combining the accuracy of ML with the interpretability of physics laws. The main contributions of this paper are:
\begin{enumerate}
    \item Magnetic core loss modeling is reformulated as an ML-based function identification problem, in the same analytical spirit as the classic SE.
    \item A sparse identification framework is proposed to automatically discover the dominant candidate functions, achieving sparsity and physical interpretability.
    \item Learnable parameters are introduced into the candidate functions to match the underlying physics and fractional power laws, leading to exceptional accuracy.
\end{enumerate}

\section{Learnable Symbolic Sparse Identification}
\label{sec:lssi}

\subsection{Problem Formulation and Candidate Library Construction}

In macroscopic magnetic characterization, the core loss $P_v$ under sinusoidal drive is traditionally assumed to be an unknown nonlinear function of the frequency $f$ and the peak flux density $B$, which can be considered as a parsimonious linear combination of distinct dominant electromagnetic loss mechanisms \cite{1041954}. Mathematically, $P_v$ is formulated as:
\begin{equation}
    P_{v}(f, B) = \boldsymbol{\Theta}(f, B; \boldsymbol{\Lambda}) \cdot \boldsymbol{\Xi}
\end{equation}
where the symbolic sparse identification library $\boldsymbol{\Theta}(f, B; \boldsymbol{\Lambda}) = [\theta_1(f, B; \boldsymbol{\Lambda}_1), \theta_2(f, B; \boldsymbol{\Lambda}_2), \dots, \theta_M(f, B; \boldsymbol{\Lambda}_M)]$ is a defined set of $M$ candidate functions; $\boldsymbol{\Xi} = [\xi_1, \xi_2, \dots, \xi_M]^T$ is a sparse vector of linear scaling coefficients; and the parameter set $\boldsymbol{\Lambda} = \left\{ \boldsymbol{\Lambda}_1, \boldsymbol{\Lambda}_2, \dots, \boldsymbol{\Lambda}_M\right\}$ represents the internal learnable nonlinear parameters (e.g., scaling exponents unique to that specific mechanism). The objective of the LSSI framework is to discover both the dominant candidate functions from the library and their precise nonlinear parameters simultaneously.

The library has significant importance on the model accuracy and interpretability. In this problem, $M=10$ specific library terms are inspired by electromagnetic physics:

\begin{figure}[t]
    \centering
    \includegraphics[width=\textwidth]{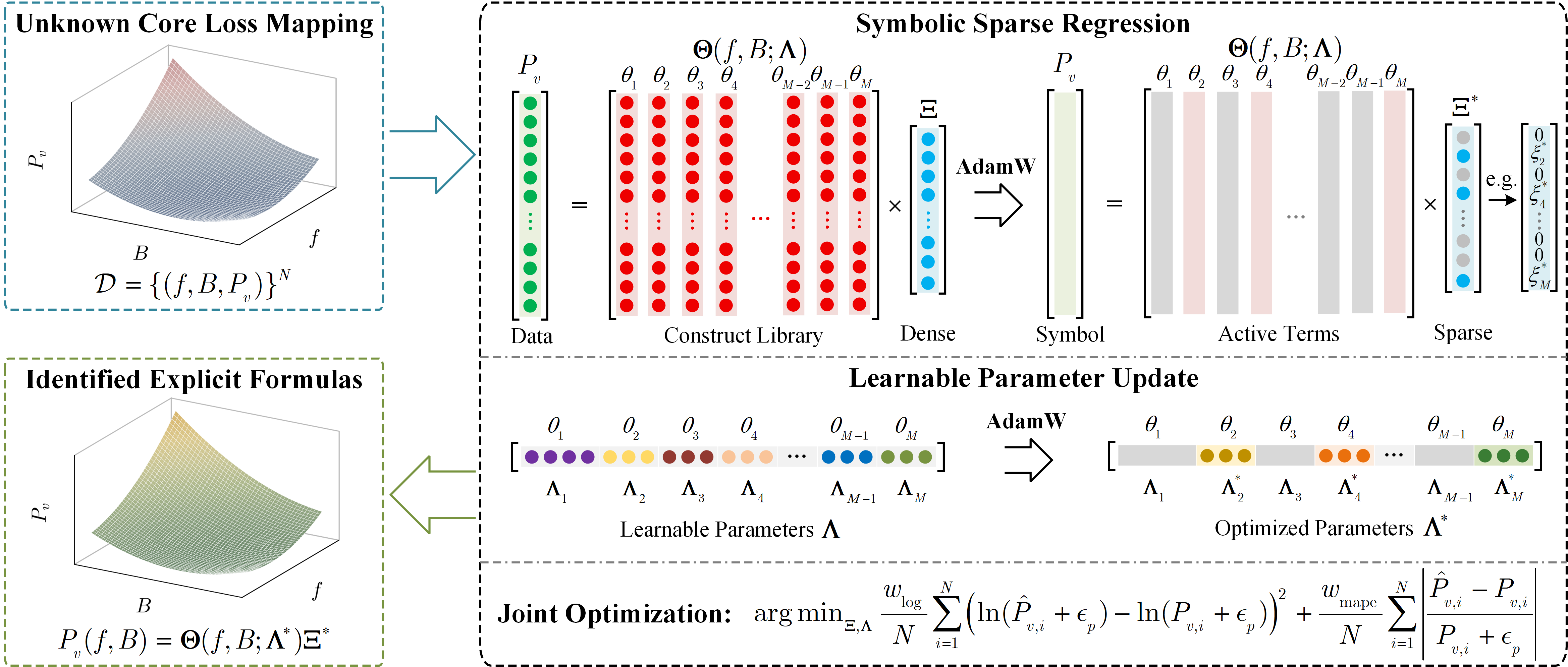}
    \caption{Overview of the proposed LSSI framework. Experimental core loss
data $\mathcal{D}=\{(f,B,P_v)\}^N$ are first normalized and fed into a
physics-inspired candidate library $\boldsymbol{\Theta}(f,B;\boldsymbol{\Lambda})$
containing hysteresis, eddy-current, anomalous, saturation, and auxiliary
terms. The sparse coefficients $\boldsymbol{\Xi}$ and the internal learnable
parameters $\boldsymbol{\Lambda}$ are then jointly optimized under the hybrid
Log-MAPE objective, with decoupled weight decay applied only to
$\boldsymbol{\Xi}$ to promote sparsity and threshold pruning removing
uninformative terms. The framework finally outputs an explicit, physically
interpretable core loss equation with only a few active terms.}
    \label{fig:LSSI}
\end{figure}


\begin{figure}[!t]
\centering
\resizebox{0.65\textwidth}{!}{%
\begin{tikzpicture}[
    x=1.1cm, y=0.95cm, font=\small,
    lat/.style   ={circle, fill=gray!42, draw=gray!60, inner sep=2pt},
    tgt/.style   ={star, star points=5, star point ratio=0.5,
                   fill=orange!85!red, draw=red!55!black, inner sep=3pt},
    seed/.style  ={circle, draw=blue!80, fill=blue!10, line width=1.1pt, inner sep=3pt},
    glide/.style ={-{Stealth[length=7pt,width=6pt]}, blue!80, line width=1.4pt},
    stack/.style ={-{Stealth[length=4pt]}, red!45, semithick},
    ax/.style    ={-{Stealth[length=8pt]}, thick},
    proj/.style  ={dashed, red!55, thin},
]
\begin{scope}[shift={(0.5,6)}]
  \node[lat]  at (0,0){};      \node[right=3pt] at (0.15,0){fixed integer term};
  \node[tgt]  at (4.5,0){};    \node[right=3pt] at (4.65,0){true fractional exponent};
  \node[seed] at (0,-0.4){};   \node[right=3pt] at (0.15,-0.4){learnable seed};
  \draw[glide] (4.5,-0.4)--(5.1,-0.4);
  \node[right=3pt] at (5.2,-0.37){exponent adaptation};
\end{scope}
\begin{scope}[shift={(0.5,0)}]
  \fill[red!8] (1,2) rectangle (2,3);
  \foreach \x in {0,1,2,3}{\draw[gray!14] (\x,0)--(\x,4.3);}
  \foreach \y in {0,1,2,3,4}{\draw[gray!14] (0,\y)--(3.3,\y);}
  \draw[ax] (-0.15,0)--(3.8,0) node[right]{$\alpha$};
  \draw[ax] (0,-0.15)--(0,4.8) node[above]{$\beta$};
  \foreach \x in {1,2,3}{\node[below] at (\x,0){\x};}
  \foreach \y in {1,2,3,4}{\node[left] at (0,\y){\y};}
  \foreach \x in {0,1,2,3}{\foreach \y in {0,1,2,3,4}{\node[lat] at (\x,\y){};}}
  \node[tgt] (a1) at (1.20,2.68){};
  \node[tgt] (a2) at (2.50,2.50){};
  \node[tgt] (a3) at (1.19,3.29){};
  \draw[red!40,dashed] (1,2) rectangle (2,3);
  \foreach \c in {(1,2),(2,2),(1,3),(2,3)}{\draw[stack] \c -- (a1);}
  \node[red!70,align=center] at (1.5,1.0){$\ge 4$ integer terms\\$\Rightarrow$ residual error};
  \node[font=\large\bfseries] at (-0.8,5.0){a)};
\end{scope}
\begin{scope}[shift={(5.2,0)}]
  \fill[blue!4] (0,0) rectangle (3.3,4.3);
  \foreach \x in {0,1,2,3}{\draw[gray!14] (\x,0)--(\x,4.3);}
  \foreach \y in {0,1,2,3,4}{\draw[gray!14] (0,\y)--(3.3,\y);}
  \draw[ax] (-0.15,0)--(3.8,0) node[right]{$\alpha$};
  \draw[ax] (0,-0.15)--(0,4.8) node[above]{$\beta$};
  \foreach \x in {1,2,3}{\node[below] at (\x,0){\x};}
  \foreach \y in {1,2,3,4}{\node[left] at (0,\y){\y};}
  \foreach \x in {0,1,2,3}{\foreach \y in {0,1,2,3,4}{\node[lat] at (\x,\y){};}}
  \node[seed] (s1) at (1,2){};
  \node[seed] (s2) at (2,2){};
  \node[seed] (s3) at (1,3){};
  \node[tgt] (b1) at (1.20,2.68){};
  \node[tgt] (b2) at (2.50,2.50){};
  \node[tgt] (b3) at (1.19,3.29){};
  \draw[glide] (s1) to[bend left=20]  (b1);
  \draw[glide] (s2) to[bend right=12] (b2);
  \draw[glide] (s3) to[bend left=14]  (b3);
  \node[above right=0pt] at (b1){$\theta_1$};
  \node[above right=1pt] at (b2){$\theta_2$};
  \node[above right=0pt] at (b3){$\theta_4$};
  \draw[proj] (b1)--(1.20,0); \draw[proj] (b1)--(0,2.68);
  \node[below,red!70] at (1.50,0.45){$1.20$};
  \node[left, red!70]  at (-0.1,2.68){$2.68$};
  \node[font=\large\bfseries] at (-0.8,5.0){b)};
\end{scope}
\end{tikzpicture}
}
\caption{Learnable symbolic identification in the power-law exponent space
$(\alpha,\beta)$ of $f^{\alpha}B^{\beta}$. (a) A fixed integer dictionary can
only approximate a fractional target by superposing several neighboring
lattice terms. (b) LSSI releases the exponents as learnable parameters, so a
single seed migrates continuously onto the true fractional exponents.}
\label{fig:learnable_lib}
\end{figure}
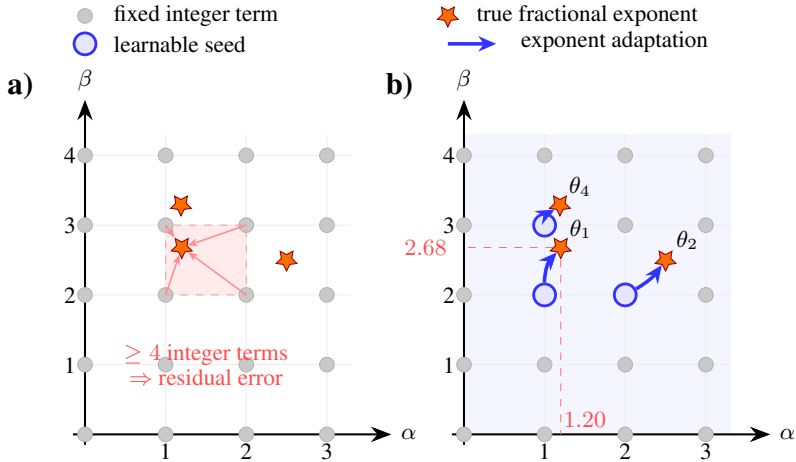


\subsubsection{Hysteresis Loss Term}
The hysteresis loss term $\theta_1(f, B; \alpha_h, \beta_h')= f^{\alpha_h} B^{\beta_h'}$ is modeled based on the Jordan and Steinmetz formulations. Specifically, $\alpha_h \approx 1$ and $\beta_h' = \beta_h + \gamma_h \ln(B + \epsilon_{h})$ is dynamically modified to model nonlinear structural shifts near saturation, where $\gamma_h$ is a learnable coupling coefficient and $\epsilon_{h}$ is a safety regularizer to prevent numerical singularity.

\subsubsection{Eddy Current Loss and Anomalous Loss Terms}
The eddy current loss term $\theta_2 (f, B; \alpha_e, \beta_e) = f^{\alpha_e} B^{\beta_e}$ is derived from Maxwell equations, where $\alpha_e \approx 2$ and $\beta_e \approx 2$; while the anomalous loss term $\theta_3 (f, B; \alpha_a, \beta_a) = f^{\alpha_a} B^{\beta_a}$ accounts for eddy currents around moving domain walls, typically exhibiting fractional exponents ($\alpha_a \approx 1.5, \beta_a \approx 1.5$).

\subsubsection{Magnetic Saturation Terms}
$\theta_4 (f, B; \alpha_s, \beta_s) = f^{\alpha_s} B^{\beta_s}$ and $\theta_5 (f, B; \delta)= f e^{\delta B}$ are an unconstrained power-law term and an exponential saturation term to flexibly map the highly nonlinear regime as a core approaches saturation and pure power-law terms fail.

\subsubsection{Auxiliary Polynomial and Bias Terms}
Cross-coupling terms ($\theta_6 (f, B)=fB$, $\theta_7 (f, B)=fB^2$), isolated linear parameters ($\theta_8(f, B)=f$, $\theta_9(f, B)=B$), and a stationary bias unit ($\theta_{10} (f, B)=1$) are also introduced to isolate experimental measurement offsets, minor unmodeled thermal drift components, and secondary physical cross-couplings, respectively.

\subsection{Learnable Parameters in Sparse Identification of Magnet Core Loss}

The rationale behind the learnable formulation becomes transparent once the
library is viewed geometrically. Most physically meaningful candidates in
$\boldsymbol{\Theta}$ share the separable form $f^{\alpha}B^{\beta}$, so that
each loss mechanism corresponds to a single point in the two-dimensional
exponent space $(\alpha,\beta)$, as depicted in
Fig.~\ref{fig:learnable_lib}.

Conventional sparse identification pins every candidate to the integer lattice
of this space, as shown in Fig.~\ref{fig:learnable_lib}(a). Since magnetic
losses intrinsically obey fractional power laws, the true exponents seldom
coincide with a lattice point, and each mechanism must instead be synthesized
as a weighted superposition of its four surrounding integer terms. Three
consequences follow: the active support of $\boldsymbol{\Xi}$ is inflated and
sparsity is lost; the one-to-one correspondence between terms and physical
mechanisms is severed, so the identified expression is no longer
interpretable; and an irreducible residual persists that no choice of
$\boldsymbol{\Xi}$ can eliminate. Refining the lattice does not resolve the
difficulty either, because densifying the dictionary inflates $M$ and renders
adjacent columns of $\boldsymbol{\Theta}$ nearly collinear, thereby
ill-conditioning the sparse regression itself.

LSSI resolves this conflict by rendering the dictionary adaptive, as
illustrated in Fig.~\ref{fig:learnable_lib}(b). Each candidate is initialized
at a physically motivated seed inherited from classical electromagnetic
theory, after which its exponents are released as learnable parameters
$\boldsymbol{\Lambda}$ and migrate continuously within the admissible domain
$\Omega_{\boldsymbol{\Lambda}}$ under gradient updates. The search is thus
transferred from an enumeration over a discrete grid to an optimization over
a continuous manifold: a candidate term is no longer approximated, but
relocated. Each mechanism is consequently captured by exactly one active term
whose fractional exponents are read off directly as continuous values, such as
$\alpha = 1.20$ and $\beta = 2.68$ in Fig.~\ref{fig:learnable_lib}(b).

This constitutes the central idea of the present work. Sparsity and
expressiveness, which compete against each other under a fixed dictionary,
become mutually compatible once the dictionary is parameterized, because
expressiveness is now supplied by the continuous parameters
$\boldsymbol{\Lambda}$ rather than by the cardinality of the active support of
$\boldsymbol{\Xi}$. This decoupling is what enables the compact four-term
expression reported in Section~3.

\subsection{Proposed Learnable Symbolic Sparse Identification}

Core losses typically span multiple orders of magnitude across wide operating ranges. To ensure uniform percentage accuracy, a hybrid Log-Mean Absolute Percentage Error (MAPE) loss function is used:
\begin{equation}
\label{eq:loss_function}
\begin{aligned}
    \mathcal{L}(\boldsymbol{\Xi}, \boldsymbol{\Lambda}) = & \frac{w_{\text{log}}}{N} \sum_{i=1}^{N} \left( \ln(\hat{P}_{v,i}(\boldsymbol{\Xi}, \boldsymbol{\Lambda}) + \epsilon_p) - \ln(P_{v,i} + \epsilon_p) \right)^2 \\
     & + \frac{w_{\text{mape}}}{N} \sum_{i=1}^{N} \left| \frac{\hat{P}_{v,i}(\boldsymbol{\Xi}, \boldsymbol{\Lambda}) - P_{v,i}}{P_{v,i} + \epsilon_p} \right|
\end{aligned}
\end{equation}
where $\epsilon_p$ is a small regularizer, $w_{\text{log}} \in \mathbb{R}^+$ and $w_{\text{mape}} \in \mathbb{R}^+$ are balancing weights, $N$ is the total number of experimental data samples, $P_{v,i}$ is the measured core loss for sample $i$, and $\hat{P}_{v,i}(\boldsymbol{\Xi}, \boldsymbol{\Lambda})$ is the prediction of the framework. The global optimization problem is then formulated as:
\begin{equation}
\min_{\boldsymbol{\Xi}, \boldsymbol{\Lambda}} \quad \mathcal{L}(\boldsymbol{\Xi}, \boldsymbol{\Lambda}), \quad
\text{s.t.} \quad \boldsymbol{\Xi} \succeq \mathbf{0}, \boldsymbol{\Lambda} \in \Omega_{\boldsymbol{\Lambda}}
\end{equation}
where $\boldsymbol{\Xi} \succeq \mathbf{0}$ ensures that active physical mechanisms strictly dissipate energy ($P_v \ge 0$). The parameters $\mathbf{\Lambda}$ are bounded by minima $\mathbf{\Lambda}_{\min}$ and maxima $\mathbf{\Lambda}_{\max}$ that define the feasible domain $\Omega_{\boldsymbol{\Lambda}}$ derived from classical electromagnetic theory.

In the discovery phase, parameters are updated by the AdamW optimization algorithm. The moment vectors $\mathbf{m}_t$ and $\mathbf{v}_t$ in AdamW tracking the gradients for both parameter spaces are computed as:
\begin{equation}
\begin{aligned}
    \mathbf{m}_t^{\boldsymbol{\Xi,\Lambda}} & = \sigma_1 \mathbf{m}_{t-1}^{\boldsymbol{\Xi,\Lambda}} + (1 - \sigma_1)\nabla_{\boldsymbol{\Xi,\Lambda}} \mathcal{L}(\boldsymbol{\Xi}_t, \boldsymbol{\Lambda}_t) \\
    \mathbf{v}_t^{\boldsymbol{\Xi,\Lambda}} & = \sigma_2 \mathbf{v}_{t-1}^{\boldsymbol{\Xi,\Lambda}} + (1 - \sigma_2)(\nabla_{\boldsymbol{\Xi,\Lambda}} \mathcal{L}(\boldsymbol{\Xi}_t, \boldsymbol{\Lambda}_t))^2
\end{aligned}
\end{equation}
where $\sigma_1, \sigma_2 \in [0, 1)$ are the first and second moment exponential decay rates, and exponents on vectors denote element-wise operations. Bias-corrections are then performed to compensate for initializations at the origin:
\begin{equation}
    \hat{\mathbf{m}}_t^{\boldsymbol{\Xi,\Lambda}} = \frac{\mathbf{m}_t^{\boldsymbol{\Xi,\Lambda}}}{1 - \sigma_1^t}, \quad \hat{\mathbf{v}}_t^{\boldsymbol{\Xi,\Lambda}} = \frac{\mathbf{v}_t^{\boldsymbol{\Xi,\Lambda}}}{1 - \sigma_2^t}
\end{equation}

The parameter update equations are executed at step $t$ using the base learning rate $\eta \in \mathbb{R}^+$. Specifically, a decoupled weight decay coefficient $\lambda_{\text{wd}} \in \mathbb{R}^+$ (i.e., $L_2$ weight penalty) is applied exclusively to $\boldsymbol{\Xi}$ to ensure its sparsity (i.e., most of its elements are zero), while $\boldsymbol{\Lambda}$ is unpenalized.
\begin{equation}
    \boldsymbol{\Xi}_{t+1} = \text{P}_{\mathcal{H}} \left[ \boldsymbol{\Xi}_t - \eta \left( \frac{\hat{\mathbf{m}}_t^{\boldsymbol{\Xi}}}{\sqrt{\hat{\mathbf{v}}_t^{\boldsymbol{\Xi}}} + \epsilon_a} + \lambda_{\text{wd}} \boldsymbol{\Xi}_t \right) \right]
\end{equation}
\begin{equation}
    \boldsymbol{\Lambda}_{t+1} = \text{P}_{\Omega_{\boldsymbol{\Lambda}}} \left[ \boldsymbol{\Lambda}_t - \eta \left( \frac{\hat{\mathbf{m}}_t^{\boldsymbol{\Lambda}}}{\sqrt{\hat{\mathbf{v}}_t^{\boldsymbol{\Lambda}}} + \epsilon_a} \right) \right]
\end{equation}
where $\epsilon_a$ is a smoothing regularizer, $\text{P}_{\mathcal{H}}[\mathbf{x}] = \max(\mathbf{0}, \mathbf{x})$ represents the hard projection operator onto the non-negative orthant constraint space, and $\text{P}_{\Omega_{\boldsymbol{\Lambda}}}[\cdot]$ denotes the element-wise clipping operator onto the explicit physical box constraint boundaries defined in the global optimization problem.

Uninformative candidate terms are automatically pruned from the library using a mask vector $\mathbf{D} \in \{0, 1\}^M$. Denote $\boldsymbol{\tau}$ as the pruning threshold vector. During AdamW, $\mathbf{D}$ is updated element-wise by the Heaviside step function $H(\cdot)$ as $\mathbf{D}_t = H(\boldsymbol{\Xi}_t - \boldsymbol{\tau})$ that outputs 0 and 1, and the model is updated by ${P}_{v}(\boldsymbol{\Xi}_t, \boldsymbol{\Lambda}_t)= \boldsymbol{\Theta}(f, B; \boldsymbol{\Lambda}_t) (\boldsymbol{\Xi}_t \odot \mathbf{D}_t)$.

\section{Case Study and Experimental Validation}

\begin{figure}[t]
    \centering
    \includegraphics[width=\textwidth]{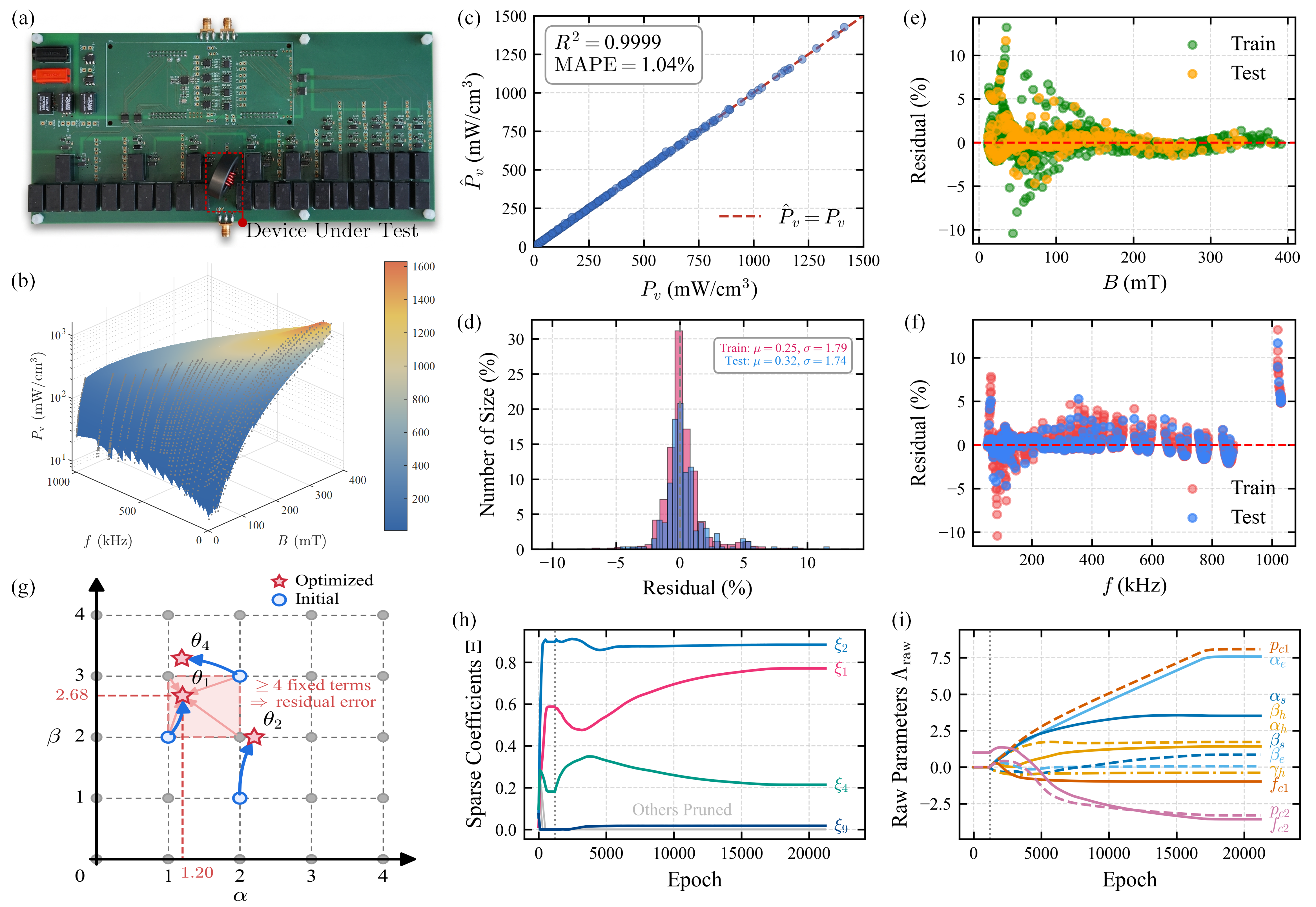}
    \caption{Experimental validation. (a) Core loss measurement system. (b) Measured core loss mapping. (c) Predicted versus measured core loss. (d) Residual distribution for the training and test sets. (e-f) Residuals as functions of $B$ and $f$. (g) Trajectory example of learnable parameters versus the combination of fixed terms. (h-i) Evolution of the sparse coefficients and raw learnable parameters.}
    \label{fig:result}
\end{figure}

The proposed LSSI framework is implemented on high-fidelity core loss datasets with sinusoidal excitations $\mathcal{D} = \{(f, B, P_{v})\}^N$. Specifically, the datasets used in this case include various frequency and flux density levels under \SI{25}{\degreeCelsius} and are collected from practical experiments on Fair-Rite 95 ferrite cores by automated parallel-resonant methods capable of sub-MHz-level acquisition, as shown in Fig.~\ref{fig:result}(a), which is considered as ground truth for the framework \cite{10977470, 11121265, wang2026resonantmethodbasedfullyautomated}. The core loss mapping is shown in Fig.~\ref{fig:result}(b). For brevity, all dataset and code used for the studied case are available in the public online repository\footnote{Source code available at: \url{https://github.com/Aaron-H-Wang/LSSI}}.

\begin{table}[t]
\centering
\caption{Predefined Hyperparameters for LSSI Training}
\label{tab:hyperparameters}
\renewcommand{\arraystretch}{1.3}
\begin{tabular}{lclc}
\hline
\rowcolor{gray!15}
\text{Parameter} & \text{Value} & \text{Parameter} & \text{Value} \\ \hline
Learning Rate ($\eta$) & $0.001$ & Weight Decay ($\lambda_{\text{wd}}$) & $0.002$ \\
AdamW Rate ($\sigma_1$) & $0.9$ & AdamW Rate ($\sigma_2$) & $0.999$ \\
Log Weight ($w_{\text{log}}$) & $0.5$ & MAPE Weight ($w_{\text{mape}}$) & $0.5$ \\
Pruning Threshold ($\tau$) & $0.001$ & All Regularizers ($\epsilon$) & $10^{-8}$ \\ \hline
\end{tabular}
\end{table}

\begin{table}[t]
\centering
\caption{Identified Parameters of the Discovered Core Loss Model}
\label{tab:params}
\renewcommand{\arraystretch}{1.3}
\begin{tabular}{lccccc}
\hline
\rowcolor{gray!15}
\text{Symbol} & \text{Value} & \text{Symbol} & \text{Value} & \text{Symbol} & \text{Value} \\
\hline
$f_{\text{norm}}$        & $373  $~kHz        & $\xi_1^{*}$       & $0.7685$     & $\alpha_s^{*}$ & $1.191$ \\
$B_{\text{norm}}$        & $103  $~mT         & $\xi_2^{*}$       & $0.8836$     & $\beta_s^{*}$  & $3.289$ \\
$P_{\text{norm}}$        & $304  $~mW/cm$^3$  & $\xi_4^{*}$       & $0.2146$     & $\alpha_e^{*}$ & $2.200$ \\
$f_{c1}^{*}$             & $1150 $~kHz        & $\xi_9^{*}$       & $0.0174$     & $\beta_e^{*}$  & $2.006$ \\
$f_{c2}^{*}$             & $1192 $~kHz        & $\gamma_h^{*}$    & $-0.387$     & $\alpha_h^{*}$ & $1.202$ \\
$p_{c1}^{*}$             & $3.999$            & $p_{c2}^{*}$      & $1.108$      & $\beta_h^{*}$  & $2.679$ \\
\hline
\end{tabular}
\end{table}

\subsection{Implementation and Results}

The framework is implemented using PyTorch with embedded differentiation for all gradient computations. The datasets including more than 1000 data points are split into a \SI{80}{\%} training set and a \SI{20}{\%} test set. All related hyperparameters are shown in Table~\ref{tab:hyperparameters}. To ensure training stability and physical plausibility, the input data are normalized to $[0, 1]$ based on the ensemble statistics to maintain consistent gradient scales across distinct physical units, and all learnable parameters $\mathbf{\Lambda}$ are passed through a logistic sigmoid mapping where $\mathbf{\Lambda}_{\text{raw}}$ are trained in the framework:
\begin{equation}
   \mathbf{\Lambda} = \mathbf{\Lambda}_{\min} + (\mathbf{\Lambda}_{\max} - \mathbf{\Lambda}_{\min}) \cdot \text{Sigmoid}(\mathbf{\Lambda}_{\text{raw}})
\end{equation}

The explicit formulas discovered by the framework mainly include $\theta_1$, $\theta_2$, $\theta_4$ and $\theta_9$, which can be expressed as:
\begin{equation}
\begin{aligned}
P_{v,n}=& \xi_1^{*} f_{n}^{\alpha_h^{*}}
   B_{n}^{\beta_h^{*} + \gamma_h^{*}\log(B_{n})} R_h(f) + \xi_2^{*} f_{n}^{\alpha_e^{*}}
   B_{n}^{\beta_e^{*}} R_e(f) \\
   &+ \xi_4^{*} f_{n}^{\alpha_s^{*}}
   B_{n}^{\beta_s^{*}} + \xi_9^{*}B_{n}
\end{aligned}
\end{equation}
where $P_{v,n}=P_v/P_{\text{norm}}, f_{n}=f/f_{\text{norm}}, B_{n}=B/B_{\text{norm}}$, $R_h$ and $R_e$ are high-frequency roll-off factors that suppress the hysteresis and eddy-current terms, respectively:
\begin{equation}
R_h(f) = \frac{1}{1 + (f/f_{c1}^{*})^{p_{c1}^{*}}}, \quad
R_e(f) = \frac{1}{1 + (f/f_{c2}^{*})^{p_{c2}^{*}}}
\end{equation}

The accuracy of the discovered model is validated on the test set. As shown in Fig.~\ref{fig:result}(c), the predicted $\hat{P}_v$ exhibits excellent agreement with the measured $P_v$ ($R^2 = 0.9999$ and $\text{MAPE} = 1.04\%$). The residual distributions in Fig.~\ref{fig:result}(d) are sharply concentrated about zero and nearly unbiased, indicating that the model generalizes without overfitting. The residuals against $f$ and $B$ in Fig.~\ref{fig:result}(e) and Fig.~\ref{fig:result}(f) further verify that no systematic physical mechanism is omitted, confirming that the physics-aware library adequately captures the dominant loss contributions. The evolution of $\mathbf{\Xi}$ and $\mathbf{\Lambda}$ throughout training is depicted in Fig.~\ref{fig:result}(h) and Fig.~\ref{fig:result}(i), respectively, where both the sparse coefficients and the learnable parameters converge to stable values well, confirming the optimization robustness. Fig.~\ref{fig:result}(g) shows the trajectories of learnable parameters in active terms and highlights that learnable parameters are flexible in discovering nonlinear equations that even a combination of multiple terms with fixed structures is incapable of, leading to exceptional sparsity and accuracy.

\begin{table}[t]
\centering
\caption{Numerical Comparison of State-of-the-art Modeling Methods.}
\label{tab:comparison}
\renewcommand{\arraystretch}{1.3}
\begin{tabular}{lccccc}
\hline
\rowcolor{gray!15}
Metrics & SE & RF & FNN & SSI & \textbf{LSSI} \\
\hline
MAPE         & $10.92\%$ & $5.13\%$ & $1.37\%$ & $1.56\%$ & $\mathbf{1.04\%}$ \\
$R^2$ Score  & $0.9868$  & $0.9961$ & $0.9996$ & $0.9997$ & $\mathbf{0.9999}$ \\
Para. Num.   & $3$       & $199{,}555$ & $4417$ & $7/103$ & $\mathbf{15/24}$ \\
Active Terms & $1$       & N/A & N/A & $7$ & $\mathbf{4}$ \\
\hline
\end{tabular}

\vspace{1mm}
\footnotesize
\raggedright
$^{*}$ SSI denotes LSSI without learnable parameters in the candidate functions;
Para. Num. = Parameter Number (all or learned / all).
\end{table}

\begin{figure}[t]
    \centering
    \includegraphics[width=0.5\textwidth]{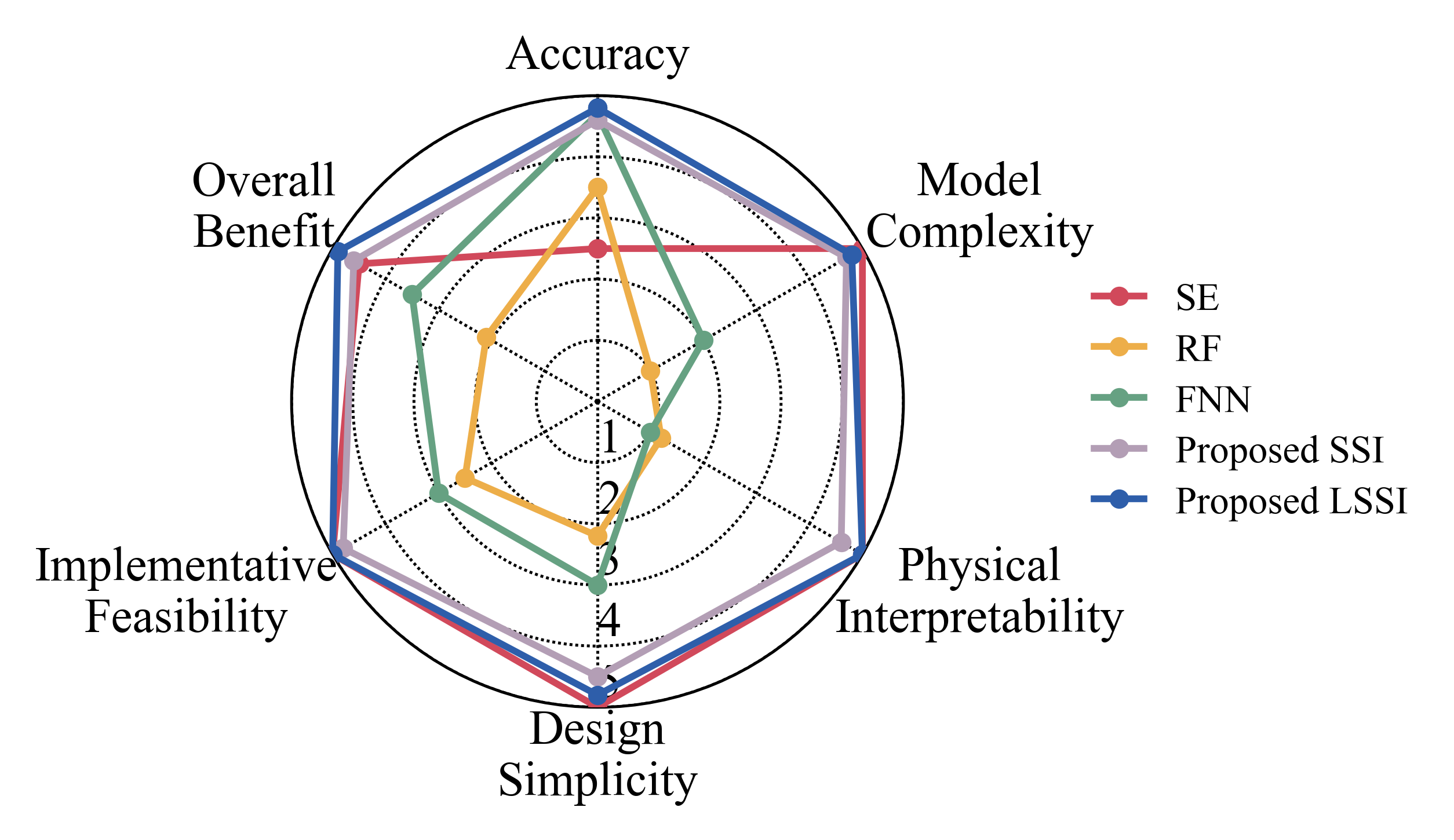}
    \caption{Performance comparison of state-of-the-art modeling methods.}
    \label{fig:radar_chart}
\end{figure}

\subsection{State-of-the-art Comparison}

To verify the superiority of the proposed framework, LSSI is compared against the empirical SE, a RF regressor, an FNN, and an SSI framework with the quantitative results summarized in Table~\ref{tab:comparison}. Notably, LSSI delivers superior accuracy using merely $15$ parameters and condenses the underlying physics into an analytical equation with only $4$ active terms, which represents significant improvement over the SE model and the data-driven baselines. The introduction of learnable parameters leads to improved accuracy and sparsity when compared to the SSI model.

A multi-dimensional performance comparison among these methods is illustrated in Fig.~\ref{fig:radar_chart}. In general, LSSI occupies the outermost envelope across accuracy, model complexity, physical interpretability, design simplicity, implementative feasibility, thereby offering the most favorable overall benefit and outperforming other state-of-the-art methods.

\section{Conclusion}
This paper proposed a Learnable Symbolic Sparse Identification (LSSI) framework that discovers explicit and physically-interpretable magnetic core loss equations directly from experimental data. LSSI formulates core loss modeling as a symbolic regression problem, establishes a sparse identification framework over a physics-based candidate library, and introduces learnable parameters to capture fractional power laws with different structures. Specifically, a hybrid Log-MAPE loss function combined with AdamW-based joint optimization and threshold pruning ensures both sparsity and plausibility across operating ranges. Consequently, LSSI is capable of simultaneously identifying the dominant loss mechanisms and their precise parameters. Experimental validation on Fair-Rite 95 ferrite data demonstrates that LSSI condenses the underlying physics into a compact analytical expression with only 4 active terms, obtaining an $R^2$ of $0.9999$ and a MAPE of $1.04\%$, outperforming other methods in accuracy while using far fewer parameters than the black-box methods. Overall, LSSI offers a highly accurate, physically transparent, and design-friendly modeling approach for high-frequency magnetics. Future work will extend the framework to temperature-dependent and non-sinusoidal excitation conditions.


\bibliographystyle{unsrt}
\bibliography{references.bib}

\end{document}